\documentclass[showkeys,showpacs,superscriptaddress]{revtex4-2}
\usepackage[caption=false]{subfig}
\usepackage{amsmath}
\usepackage{amssymb}
\usepackage{graphicx}
\usepackage{bm}
\usepackage{float}
\usepackage{ulem,xcolor}
\usepackage{color,soul}
\begin{document}

\title{Mass gap for a monopole interacting with a nonlinear spinor field
}

\author{
Vladimir Dzhunushaliev
}
\email{v.dzhunushaliev@gmail.com}
\affiliation{
Institute of Nuclear Physics, Almaty 050032, Kazakhstan
}

\affiliation{
Department of Theoretical and Nuclear Physics,  Al-Farabi Kazakh National University, Almaty 050040, Kazakhstan
}
\affiliation{
Academician J.~Jeenbaev Institute of Physics of the NAS of the Kyrgyz Republic, 265 a, Chui Street, Bishkek 720071, Kyrgyzstan
}

\author{
Nassurlla Burtebayev
}
\affiliation{
Institute of Nuclear Physics, Almaty 050032, Kazakhstan
}

\author{Vladimir Folomeev}
\email{vfolomeev@mail.ru}
\affiliation{
Institute of Nuclear Physics, Almaty 050032, Kazakhstan
}
\affiliation{
Academician J.~Jeenbaev Institute of Physics of the NAS of the Kyrgyz Republic, 265 a, Chui Street, Bishkek 720071, Kyrgyzstan
}
\affiliation{
International Laboratory for Theoretical Cosmology, Tomsk State University of Control Systems and Radioelectronics (TUSUR),
Tomsk 634050, Russia
}

\author{
Jutta Kunz
}
\email{
jutta.kunz@uni-oldenburg.de
}
\affiliation{
Institut f\"ur Physik, Universit\"at Oldenburg, Postfach 2503
D-26111 Oldenburg, Germany
}

\author{Albina Serikbolova}
\email{albeni_23_95@mail.ru}
\affiliation{
Department of Theoretical and Nuclear Physics,  Al-Farabi Kazakh National University, Almaty 050040, Kazakhstan
}

\author{
Abylaikhan Tlemisov
}
\email{tlemissov-ozzy@mail.ru}
\affiliation{
Department of Theoretical and Nuclear Physics,  Al-Farabi Kazakh National University, Almaty 050040, Kazakhstan
}

%\date{\today}

\begin{abstract}
Within SU(2) Yang-Mills theory with a source of the non-Abelian gauge field in the form of a classical spinor field, 
we study the dependence of the mass gap on the coupling constant between the gauge and nonlinear spinor fields. 
It is shown that the total dimensionless energy of the monopole interacting with the nonlinear spinor fields depends 
only on the dimensionless coupling constant.
\end{abstract}


\date{\today}

\maketitle

\section{Introduction}

The presence of a mass gap in the energy spectrum of a particle-like solution in classical field theory is a rather rare phenomenon. Perhaps this property of the energy spectrum was first discovered for particle-like solutions of the nonlinear Dirac equation (see Refs.~\cite{Finkelstein:1951zz,Finkelstein:1956}). Interest in studying such a Dirac equation has emerged in the 1950's when W.~Heisenberg tried to employ this equation as a fundamental equation suitable for describing the properties of an electron~\cite{hei66}. Later, equations of this type have been used as a complicated effective theory of nucleons and mesons constructed from interacting Dirac fermions with chiral symmetry~\cite{Nambu:1961tp,Nambu:1961fr}. In Ref.~\cite{Hatsuda:1994pi}, the Nambu-Jona-Lasinio approach is considered as a low-energy effective theory of QCD.

The monopole solution in SU(2) Yang-Mills theory with a source in the form of a spinor field described by the nonlinear Dirac equation has been obtained in Ref.~\cite{Dzhunushaliev:2020qwf}. That solution differs in principle from the 't~Hooft-Polyakov monopole by the fact that it is topologically trivial, and the asymptotic behavior of the radial magnetic field is different.
It should be emphasised here that the asymptotic behavior does not permit us to introduce the notion of a magnetic charge,
since the integral of the magnetic field over a closed surface goes to zero at infinity. On the other hand,
the asymptotic behavior of the radial magnetic field is similar to the asymptotic behavior of a Maxwellian dipole magnetic field,
but the energy density is spherically symmetric, in contrast to a Maxwellian dipole.

An interesting feature of the aforementioned solution is that the energy spectrum has a global minimum~-- the mass gap.
This property of the energy spectrum is caused by the presence of the nonlinear spinor field. The reason is that, as was demonstrated earlier
in Refs.~\cite{Finkelstein:1951zz,Finkelstein:1956}, the energy spectrum of the nonlinear Dirac equation has a global minimum, which the authors referred to as  ``the lightest stable particle.''

In the present paper we continue investigations in this direction by studying the dependence of the dimensionless mass gap of the monopole  on the dimensionless coupling constant between gauge and spinor fields.

\section{Equations and Ans\"{a}tze of Yang-Mills fields coupled to a nonlinear Dirac field}
\label{YM_Dirac_scalar}

In this section we closely follow Ref.~\cite{Dzhunushaliev:2020qwf}. The Lagrangian describing a system consisting of a non-Abelian SU(2) field $A^a_\mu$ interacting with nonlinear spinor field $\psi$ can be taken in the form
\begin{equation}
\begin{split}
	\mathcal L = & - \frac{1}{4} F^a_{\mu \nu} F^{a \mu \nu}
	+ i \hbar c \bar \psi \gamma^\mu D_\mu \psi  -
	m_f c^2 \bar \psi \psi+
	\frac{\Lambda}{2} g \hbar c \left( \bar \psi \psi \right)^2.
\label{1_10}
\end{split}
\end{equation}
Here $m_f$ is the mass of the spinor field;
$
	D_\mu = \partial_\mu - i \frac{g}{2} \sigma^a
A^a_\mu
$ is the gauge-covariant derivative, where $g$ is the coupling constant and $\sigma^a$ are the SU(2) generators (the Pauli matrices);
$
	F^a_{\mu \nu} = \partial_\mu A^a_\nu - \partial_\nu A^a_\mu +
g \epsilon_{a b c} A^b_\mu A^c_\nu
$ is the field strength tensor for the SU(2) field, where $\epsilon_{a b c}$ (the completely antisymmetric Levi-Civita symbol)
are the SU(2) structure constants;  $\Lambda$ is a constant; $\gamma^\mu$ are the Dirac matrices in the standard representation;
$a,b,c=1,2,3$ are color indices and $\mu, \nu = 0, 1, 2, 3$ are spacetime indices.

Using the Lagrangian~\eqref{1_10}, one can find the corresponding field equations
\begin{eqnarray}
	D_\nu F^{a \mu \nu} &=& \frac{g \hbar c}{2}
	\bar \psi \gamma^\mu \sigma^a \psi ,
	\label{1_20}\\
	i \hbar \gamma^\mu D_\mu \psi  - m_f c \psi + \Lambda g \hbar \psi
	\left(
		\bar \psi \psi
	\right)&=& 0.
\label{1_30}
\end{eqnarray}

Our purpose is to study monopole-like solutions of these equations. To do this, we use the standard SU(2) monopole {\it Ansatz}
\begin{eqnarray}
	A^a_i &=&  \frac{1}{g} \left[ 1 - f(r) \right]
	\begin{pmatrix}
		0 & \phantom{-}\sin \varphi &  \sin \theta \cos \theta \cos \varphi \\
		0 & -\cos \varphi &   \sin \theta \cos \theta \sin \varphi \\
		0 & 0 & - \sin^2 \theta
	\end{pmatrix} , \quad
%\nonumber \\
%	&&
	i = r, \theta, \varphi  \text{ (in polar coordinates)},
\label{2_10}\\
	A^a_t &=& 0 ,
\label{2-13}
\end{eqnarray}
and the {\it Ansatz} for the spinor field from Refs.~\cite{Li:1982gf,Li:1985gf}
\begin{equation}
	\psi^T = \frac{e^{-i \frac{E t}{\hbar}}}{g r \sqrt{2}}
	\begin{Bmatrix}
		\begin{pmatrix}
			0 \\ - u \\
			\end{pmatrix},
			\begin{pmatrix}
			u \\ 0 \\
			\end{pmatrix},
			\begin{pmatrix}
			i v \sin \theta e^{- i \varphi} \\ - i v \cos \theta \\
			\end{pmatrix},
			\begin{pmatrix}
			- i v \cos \theta \\ - i v \sin \theta e^{i \varphi} \\
		\end{pmatrix}
	\end{Bmatrix},
\label{2_20}
\end{equation}
where $E/\hbar$ is the spinor frequency and the functions $u$ and $v$ depend on the radial coordinate $r$ only.
%In Eq.~\eqref{2_20}, each row describes a spin-$1/2$ fermion, and these two fermions have the same mass $m_f$ and opposite spins and are located at one point. Aside from this, for each of such fermions, the energy-momentum tensors will not be spherically symmetric (due to the existence of nondiagonal components), but their sum will give a tensor compatible with spherical symmetry of the system under consideration.

Equations for the unknown functions $f, u$, and $v$ can be obtained  by substituting the expressions~\eqref{2_10}-\eqref{2_20} into the field equations~\eqref{1_20} and \eqref{1_30},
\begin{eqnarray}
	- f^{\prime \prime} + \frac{f \left( f^2 - 1 \right) }{x^2} +
	\tilde g^2_{\Lambda} \frac{\tilde u \tilde v}{x} &=& 0 ,
\label{2_30}\\
	\tilde v' + \frac{f \tilde v}{x} &=& \tilde u \left(
		- 1+ \tilde E +
		\frac{\tilde u^2 - \tilde v^2}{x^2}
	\right) ,
\label{2_40}\\
	\tilde u' - \frac{f \tilde u}{x} &=& \tilde v \left(
		- 1 - \tilde E +
	\frac{\tilde u^2 - \tilde v^2}{x^2}
	\right).
\label{2_50}
\end{eqnarray}
Here, for convenience of making numerical calculations, we have introduced the following dimensionless variables:
$x = r/\lambda_c$,
$
	\tilde u=u\sqrt{\Lambda/\lambda_c g},
	\tilde v = v\sqrt{\Lambda/\lambda_c g},
	\tilde E = \lambda_c E/(\hbar c),
	\tilde g^2_{\Lambda} = g \hbar c\lambda_c^2/\Lambda
$, where $\lambda_c= \hbar / (m_f c)$ is the Compton wavelength. The prime denotes differentiation with respect to  $x$.

The total energy density of the monopole under consideration is
\begin{equation}
	\tilde \epsilon =
	\tilde{\epsilon}_m + \tilde \epsilon_s =\frac{1}{\tilde g^2_\Lambda}
	\left[
		\frac{{f'}^2}{ x^2} +
		\frac{\left( f^2 - 1 \right)^2}{2 x^4}
	\right] +
	\left[
		\tilde E \frac{\tilde u^2 + \tilde v^2}{x^2} +
		\frac{\left(\tilde u^2 - \tilde v^2 \right)^2}{2 x^4}
	\right].
\label{2_60}
\end{equation}
%\textcolor{red}{where the dimensionless  $\tilde\Lambda=\left(g/\lambda_c^2\right)\Lambda$}.
Here the expressions in the square brackets correspond to the dimensionless energy densities of the non-Abelian gauge fields,
$
	\tilde{\epsilon}_m \equiv
	\frac{\lambda_c^4 g^2}{\tilde g^2_\Lambda} \epsilon_m
$,
and of the spinor field,
$
	\tilde{\epsilon}_s \equiv \frac{\lambda_c^4 g^2}{\tilde g^2_\Lambda} \epsilon_s
$.

Correspondingly, the total energy of the monopole is calculated using the formula
\begin{equation}
	\tilde W_t \equiv \frac{\lambda_c g^2}{\tilde g^2_\Lambda} W_t =
	4 \pi
	\int\limits_0^\infty x^2 \tilde \epsilon d x
	= \left( \tilde{W}_t \right)_m + \left( \tilde{W}_t \right)_{s},
\label{2_70}
\end{equation}
where the energy density $\tilde \epsilon$ is taken from Eq.~\eqref{2_60}.  In the right-hand side of Eq.~\eqref{2_70}, the first term is the energy of the magnetic field and the second term is the energy of the nonlinear spinor field.

\section{Mass gap and its properties}

The purpose of this section is to study in more detail the properties of the monopole solution found in Ref.~\cite{Dzhunushaliev:2020qwf}. To understand how the solution and hence the energy of the monopole depend on the parameters of the system, let us consider the series expansion of the functions $f(x), \tilde u(x)$, and $\tilde v(x)$ appearing in Eqs.~\eqref{2_30}-\eqref{2_50} in the vicinity of the origin of coordinates:
%\begin{equation}
$$f = 1 + \frac{f_2}{2} x^2 + \ldots ,\quad
	\tilde u = \tilde u_1 x + \frac{\tilde u_3}{3!} x^3 + \ldots ,\quad
	\tilde v = \frac{\tilde v_2}{2} x^2 + \frac{\tilde v_4}{4!} x^4 + \ldots$$
%\label{T_series}
%\end{equation}
These expansions and Eqs.~\eqref{2_30}-\eqref{2_50} contain the following set of parameters:
$
	f_2, \tilde u_1, \tilde v_2, \tilde g_\Lambda, \tilde E.
$
The value of the parameter $\tilde v_2 $ can be found from Eqs.~\eqref{2_30}-\eqref{2_50},
$
	\tilde v_2 = 2 \tilde u_1 \left(
		\tilde E - 1 + \tilde\Lambda \tilde u_1^2
	\right)/3
$.
Next, for arbitrary values of $f_2$ and $\tilde u_1$, the solution of the set of equations~\eqref{2_30}-\eqref{2_50} is singular in the sense
that all functions are singular and hence the total energy of the monopole is infinite. For this reason, the set of equations~\eqref{2_30}-\eqref{2_50}
should be solved as a nonlinear eigenvalue problem for the eigenvalues
$f_2, \tilde u_1$ and eigenfunctions $f(x), \tilde u(x)$, and $\tilde v(x)$. To find the global minimum of the energy spectrum, we must construct the dependence of the total energy of the monopole on the quantity
$0~\leqslant~\tilde E~\leqslant~1$.
%For the sake of simplicity, we choose $r_0 = \hbar/(m_f c)$; then $\tilde m_f = 1$.
We therefore reach the conclusion that the mass gap of the monopole interacting with the nonlinear spinor field depends on one parameter  $\tilde g_{\Lambda}$ only. Physically, this means that the dimensionless energy of the monopole and hence its mass gap depend only on the ratio of the coupling constant $g$ between the gauge and spinor fields to the constant $\Lambda$ of the nonlinear spinor field.

\begin{figure}[H]
\begin{minipage}[t]{.45\linewidth}
	\begin{center}
		\includegraphics[width=1\linewidth]{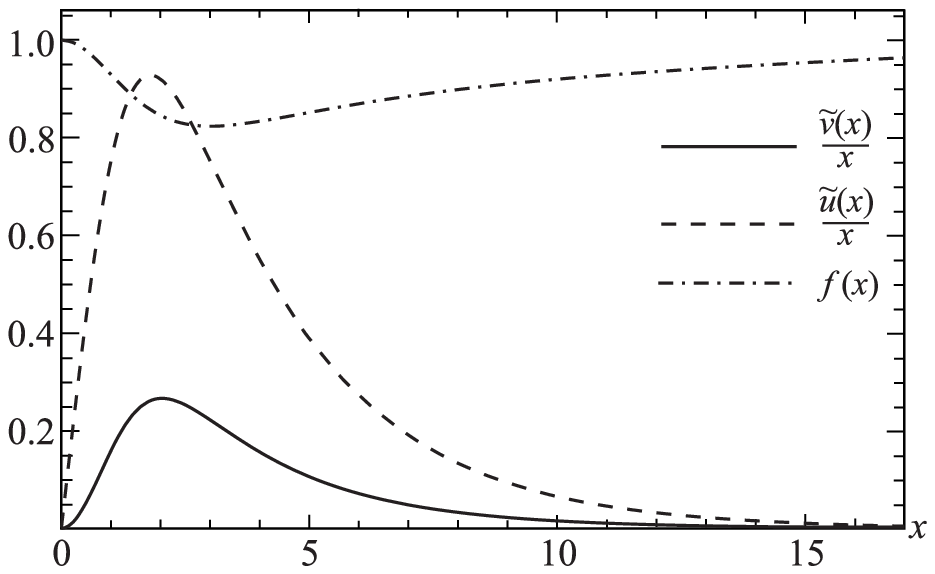}
	\end{center}
\vspace{-0.5cm}
\caption{The functions $f, \tilde u$, and $\tilde v$ for
$
	\tilde g_{\Lambda} = 1.03279, \tilde E = 0.933,
	f_2 = - 0.181365, u_1 = 0.9065608
$.
}
\label{potentials}
\end{minipage} \hfill
\begin{minipage}[t]{.47\linewidth}
	\begin{center}
		\includegraphics[width=1\linewidth]{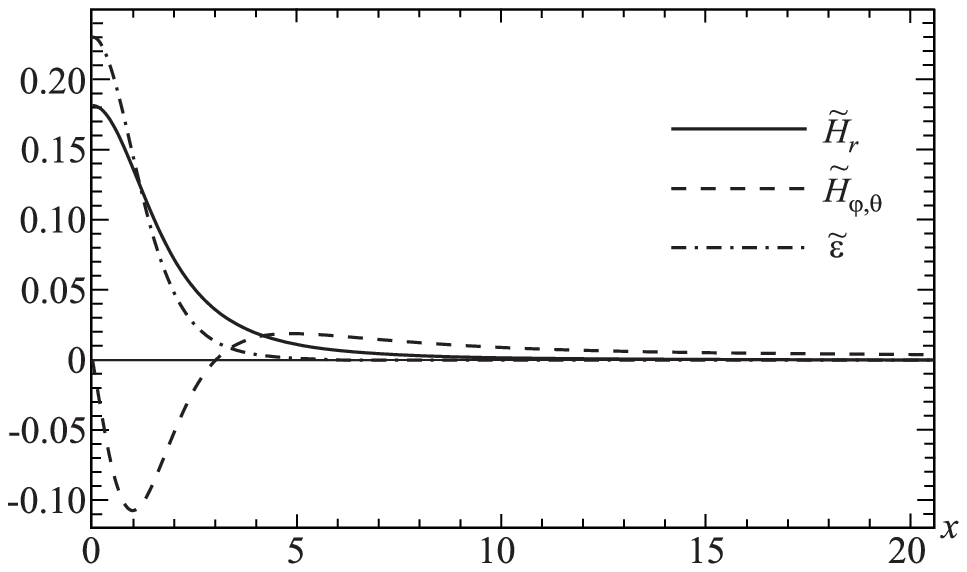}
	\end{center}
\vspace{-0.5cm}
\caption{
The magnetic fields $\tilde H_r, \tilde H_{\varphi, \theta}$ given by Eqs.~\eqref{3_10}-\eqref{3_30} and the energy density $\tilde \epsilon$ from \eqref{2_60}.
}
\label{fields}
\end{minipage} \hfill
\end{figure}

To understand the behavior of the potentials $A^a_\mu$, the field intensities $H^a_{r, \theta, \varphi}$ created by the monopole,
 and the dimensionless energy density $\tilde \epsilon$, we have plotted the graphs of these functions in
 Figs.~\ref{potentials} and \ref{fields}. The physical components of the color magnetic fields $H^a_i$ are defined as
$
H^a_i=-(1/2)\sqrt{\gamma}\,\epsilon_{i j k} F^{a j k},
$
where $i, j, k$ are space indices and $\gamma$ is the determinant of the spatial metric. This gives the components
\begin{align}
		H^a_r \sim & \frac{1 - f^2}{g r^2},
\label{3_10}\\
	H^a_{\theta} \sim &\frac{1}{g}f^{\prime},
\label{3_20}\\
	H^b_{\varphi} \sim &\frac{1}{g}f^{\prime},
\label{3_30}
\end{align}
where $a=1,2,3$ and we have dropped the dependence on the angular variables.

The results of calculations are given in Tables~\ref{energy_spestrum_1}-\ref{mass_gap}. According to these results, one can also draw the conclusion that the magnitude
of $\left( \tilde E\right)_{MG}$ at which the energy spectrum reaches its global minimum practically does not depend on the coupling constant $\tilde g_{\Lambda}$.

\begin{figure}[H]
\begin{minipage}[t]{.45\linewidth}
\begin{center}
		\includegraphics[width=1.\linewidth]{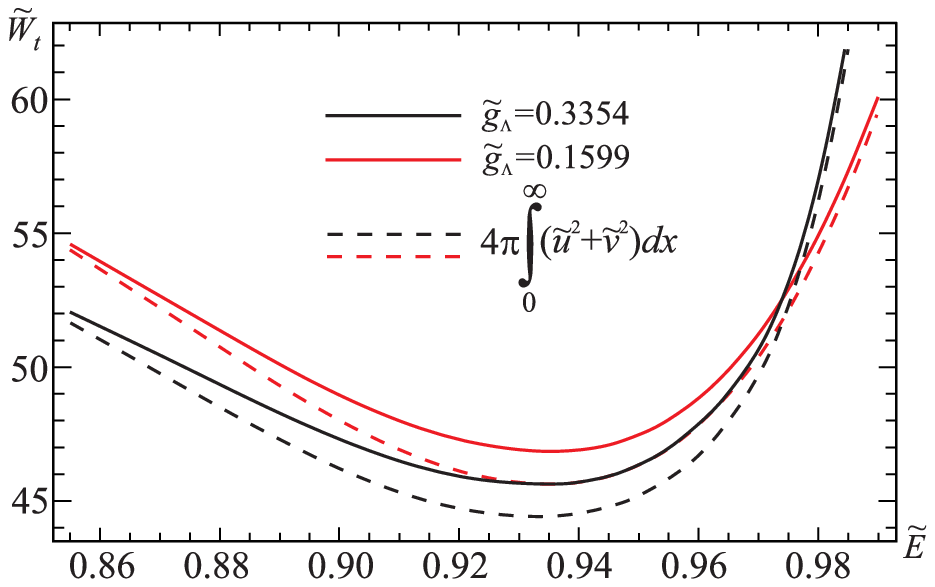}
\end{center}
\vspace{-0.5cm}
\caption{
	The energy spectrum and normalization integral for two values of the dimensionless coupling constant $\tilde g_\Lambda$.
}
\label{EnergySpectrum}
\end{minipage} \hfill
\begin{minipage}[t]{.48\linewidth}
\begin{center}
		\includegraphics[width=1.\linewidth]{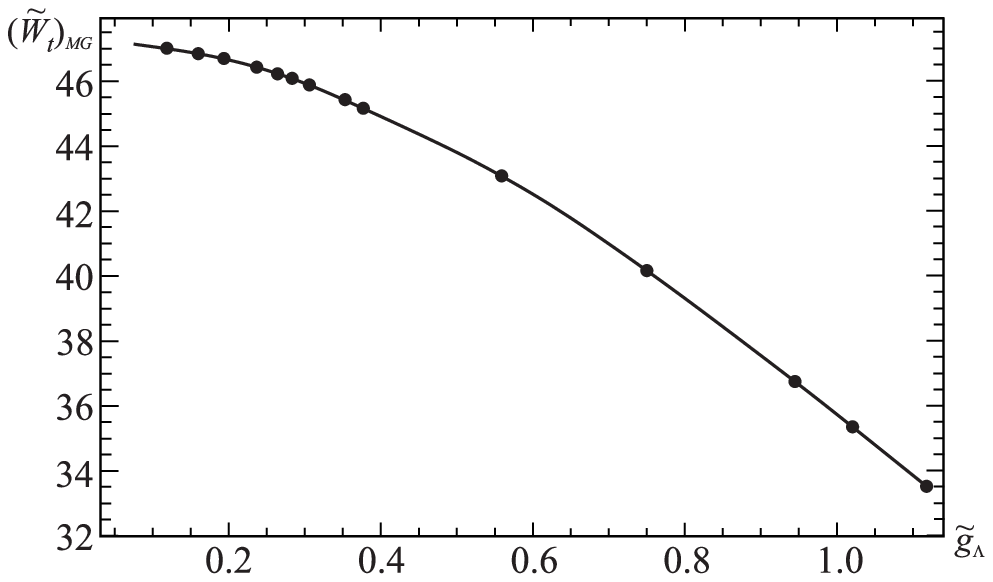}
\end{center}
\vspace{-0.5cm}
\caption{The dependence of the mass gap $\left( \tilde W_t\right)_{MG}$ on the coupling constant
 $\tilde g_\Lambda$.
}
\label{MG_location}
\end{minipage} \hfill
\end{figure}

In order to find the mass gap, we numerically solved Eqs.~\eqref{2_30}-\eqref{2_50} and found the energy spectrum for  values of $\tilde{g}_\Lambda$ lying in the range $0.075 \leqslant \tilde g_{\Lambda} \leqslant 1.02062$. In doing so, for every value of the dimensionless constant $\tilde{g}_\Lambda$, we constructed the profile of the dependence of $\tilde W_t$ on $\tilde E$. As an example of these calculations, in Tables~\ref{energy_spestrum_1} and \ref{energy_spestrum_2}, we show the calculated values of the total energy $\tilde W_t(\tilde E)$ and of the eigenvalues  $f_2$ and $\tilde u_1$ for given values of $\tilde g_\Lambda$. The profiles of these two energy spectra are shown in Fig.~\ref{EnergySpectrum}, where the normalization integral
$
	\int \psi^\dagger \psi dV
$ is also given; its physical meaning is a different matter, and we will discuss it in Sec. \ref{conclusions}.

\begin{table}[t]
\begin{minipage}[t]{.45\linewidth}
	\begin{tabular}{ |c|c|c|c|c| }
		\hline
		$\tilde{E}$ & $f_2$& $\tilde{u}_1$&$\tilde{W}_t$ \\
		\hline
		0.988 & -0.0041&0.455151&67.09535 \\
		\hline
		0.977 & -0.0076851&0.607707&54.4657\\
		\hline
		0.966 & -0.011237&0.716816&49.337095\\
		\hline
		0.955 & -0.011237&0.801936&46.983028\\
		\hline
		0.944 & -0.018214&0.871516&45.92847\\
		\hline
		0.933 & -0.021675&0.930108&45.645555\\
		\hline
		0.922 & -0.025131&0.980471&45.838035\\
		\hline
		0.855 & -0.046802&1.18361&52.05412\\
		\hline
	\end{tabular}
\caption{Eigenvalues $f_2, \tilde{u}_1$ and the total energy $\tilde{W}_t$ for $\tilde{g}_{\Lambda}= 0.3354$.}
\label{energy_spestrum_1}
%\end{table}
\end{minipage} \hfill
\begin{minipage}[t]{.45\linewidth}
%\begin{table}
	\begin{tabular}{ |c|c|c|c|c| }
		\hline
		$\tilde{E}$ & $f_2$& $\tilde{u}_1$&$\tilde{W}_t$ \\
		\hline
		0.966 & -0.0025996&0.71939&50.11534\\
		\hline
		0.944 & -0.004196&0.873927&47.03193\\
		\hline
		0.933 & -0.005001&0.932320&46.85765\\
		\hline
		0.922 & -0.005792&0.982419&47.179531\\
		\hline
		0.855 & -0.010797&1.183101&54.592154\\
		\hline
	\end{tabular}
\caption{
Eigenvalues $f_2, \tilde{u}_1$ and the total energy $\tilde{W}_t$ for $\tilde{g}_{\Lambda}= 0.1599$.
}
\label{energy_spestrum_2}
\end{minipage}
\end{table}

\begin{table}[t]
\begin{tabular}{ |c|c|c|c|c||c|c|c|c|c| }
\hline
	$\tilde{E}$ & $f_2$& $\tilde{u}_1$&$\tilde{g}_\Lambda$&$(\tilde{W}_t)_{\mathrm{Min}}$&$\tilde{E}$&$f_2$&$\tilde{u}_1$&$\tilde{g}_{\Lambda}$&$(\tilde{W}_t)_{\text{Min}}$\\ \hline
	0.89512& -0.2747&1.06415&1.02062&35.3559&0.93493&-0.01758&0.9212&0.30618&45.88048\\ \hline
   0.90190& -0.2246&1.04327&0.9449&36.7500&0.93351&-0.01541&0.9286&0.2834&46.06546\\	\hline
	0.91076& -0.1346&1.016385&0.75&40.1651&0.93538&-0.01314&0.9195&0.26516&46.20792\\	\hline
   0.92543& -0.0649&0.96094&0.559&43.0803&0.93454&-0.01068&0.9241&0.23717&46.37691\\	\hline
	0.91937 & -0.04554&0.98959&0.447&44.6403&0.93448&-0.00713&0.9248&0.19364&46.68651\\ \hline
	0.92096 & -0.03739&0.98365&0.408&45.0710&0.93559&-0.00481&0.9195&0.1599&46.824960\\	\hline
   0.92209& -0.03154&0.9796&0.377&45.3804&0.93573&-0.00265&0.9191&0.11858&46.98536\\	\hline
	0.92294&-0.02742&0.9763&0.353&45.62255&0.93547&-0.001311&0.9207&0.075&47.02922\\	\hline
\end{tabular}
\caption{Eigenvalues $f_2, \tilde{u}_1$ and the energy of the mass gap $(\tilde{W}_t)_{MG}\equiv (\tilde{W}_t)_{\text{Min}}$ as functions of the dimensionless coupling constant
$\tilde{g}_\Lambda$.
}
\label{mass_gap}
\end{table}

Next, for every spectrum, we have calculated the magnitude of the mass gap
$\left( \tilde W_t\right)_{MG} $ and plotted its dependence on $\tilde g_\Lambda$
 in Fig.~\ref{MG_location}. The corresponding values of the mass gap
 $\left( \tilde W_t\right)_{MG}$, the eigenvalues
$f_2, \tilde u_1$, and the values of the dimensionless coupling constant
$\tilde g_\Lambda$ are collected in Table~\ref{mass_gap}.

\section{Qualitative explanation of the appearance of the mass gap
}

The appearance of the mass gap is a fairly surprising phenomenon in an energy spectrum of any solution in field theories. For example,
there is no mass gap in the energy spectrum of the 't~Hooft-Polyakov monopole. It is therefore of great interest to understand
the reason for the appearance of such a phenomenon in field theory.

The set of equations \eqref{2_30}-\eqref{2_50} has regular, finite energy solutions only in the presence of the spinor field;
when it is absent, there are only trivial solutions
$
	\tilde u = \tilde v = 0,
	f = 0, \pm 1
$. On the other hand, the nonlinear Dirac equations \eqref{2_40} and \eqref{2_50}, as was demonstrated in Refs.~\cite{Finkelstein:1951zz,Finkelstein:1956},
possess regular solutions with finite energy and mass gap. It is therefore natural to conclude that the reason for the appearance of the mass gap
in the monopole solution studied here is the presence of the nonlinear spinor field described by Eqs.~\eqref{2_40} and \eqref{2_50}.
The spinor field by itself already has the mass gap. So the behavior of the solutions known for the spinor fields is only modified by the gauge field, and the presence of the spinor fields
is of fundamental importance
(see Refs.~\cite{Finkelstein:1951zz,Finkelstein:1956,Soler:1970xp}).

Fig.~\ref{EnergySpectrum} shows a typical energy spectrum $\tilde{W}_t(\tilde E)$, and one can see that as
$
	\tilde E \rightarrow  1
$  the total energy of the monopole  $\tilde W_t \rightarrow \infty$. The numerical calculations indicate that for $
	\tilde E \rightarrow  0
$ the total energy $\tilde W_t \rightarrow \infty$ as well.
What is the reason for that? To clarify this question,
let us estimate the linear size of the monopole. To do this, it is necessary to find an asymptotic behavior of the solution obtained here.
It is easy to see that asymptotically the fields behave as follows:
%\begin{equation}
$$f	(x) \approx 1 - \frac{f_\infty}{x} , \quad
	\tilde u(x) \approx \tilde u_\infty
	e^{- x \sqrt{1 - \tilde E^2}} ,\quad
	\tilde v(x) \approx
	\tilde v_\infty e^{- x \sqrt{1- \tilde E^2}} ,$$
%\label{4_10}
%\end{equation}
where $f_\infty, \tilde u_\infty$,  and $\tilde v_\infty$ are integration constants.
So the size of the monopole can be estimated as
%\begin{equation}
$$	x_{mp} \approx \frac{1}{\sqrt{1- \tilde E^2}} .$$
%\label{4_20}
%\end{equation}
This means that for $\tilde E \rightarrow 1$ the quantity $x_{mp} \rightarrow \infty$,
but the numerical solution to the system \eqref{2_30}-\eqref{2_50} indicates that in this case the functions
$\tilde v(x), \tilde u(x) \rightarrow 0$ and $f(x) \rightarrow 1$. The energy $\tilde W_t$, according to \eqref{2_60},
can be estimated as follows:
%\begin{equation}
$$	\tilde W_t \approx \frac{4}{3} \pi x^3_{mp} \tilde \epsilon(x_{\text{max}}) \approx
	\frac{2}{3}\frac{\tilde \epsilon(x_{\text{max}})}
	{\left( 1 - \tilde E\right)^{3/2}},$$
%\label{4-30}
%\end{equation}
where $x_{\text{max}}$ is the point where the energy density has a maximum. To ensure that $\tilde W_t \rightarrow \infty$, it is therefore necessary that the energy density would decrease as
%\begin{equation}
$$	\tilde \epsilon(x_{\text{max}}) \approx \left(
		1 - \tilde E
	\right)^{\frac{3}{2} - \delta},$$
%\label{4-40}
%\end{equation}
where the exponent $\delta > 0$.

For $\tilde E \rightarrow 0$, the dimensionless size of the monopole is estimated as
%\begin{equation}
$	x_{mp} \approx 1.$
%\label{4_20}
%\end{equation}
That means that the size of the monopole remains constant when $\tilde E \rightarrow 0$.
Therefore, in order for the energy of the monopole to tend to infinity, it is necessary that
$\tilde \epsilon(x_{\text{max}}) \rightarrow \infty$; this is actually observed in numerical calculations.

\section{Conclusions and discussion}
\label{conclusions}

Summarizing the results obtained in the present paper,
\begin{itemize}
\item We have shown that the dimensionless total energy $\tilde W_t $ of the monopole constructed within SU(2) Yang-Mills theory involving
two nonlinear spinor fields depends only on the dimensionless coupling constant $\tilde{g}_\Lambda$ relating the non-Abelian and spinor fields.
\item For the values of the coupling constant lying in the range
$0.075 \leqslant \tilde{g}_\Lambda \leqslant 1.02062$, spectra of the energy  $\tilde W_t(\tilde E)$ have been constructed.
For every spectrum, a minimum value, $\left( \tilde W_t\right)_{MG} \left( \tilde{g}_\Lambda\right)$, referred to as a mass gap,
has been calculated and the corresponding profile has been constructed.
%\item \textcolor{red}{\sout{In the range $0.075 \leqslant \tilde{g}_\Lambda \leqslant 1.02062$, the profile of the dependence of the mass gap on the dimensionless
%coupling constant between gauge and spinor fields has been constructed. }}
\item The qualitative analysis of the behavior of linear sizes of the monopole for $\tilde E \rightarrow 0, 1$, as well as of the corresponding energy densities,
has been carried out.
\end{itemize}

In the present paper, we have studied a monopole for which a classical nonlinear spinor field is a source of a non-Abelian gauge field. In this connection, the following questions arise: (a)~the physical meaning of the nonlinear spinor field; (b)~what an observer sees at infinity~-- a monopole
or something else; (c)~whether it is possible to consider the normalization integral as a quantity associated with the fermion number?

In response it may be pointed out that
\vspace{-0.2cm}
\begin{enumerate}
\itemsep=-0.2pt
\item[(a)] %1.
 Since, apparently, there are no classical spinor fields in nature, the nonlinear spinor field used by us should be quantized,  thus allowing only for a discrete set of bound states. While the field theory employed here is nonrenormalizable, there is an alternative point of view
(see Ref.~\cite{Dzhunushaliev:2019ham}) according to which the nonlinear spinor interaction represents some approximate phenomenological description of nonlinear interaction of the type
$
	\hat{\bar \psi} \gamma^\mu \lambda_a A^a_\mu \hat \psi
$ between spinor and gauge fields.
\item[(b)]
According to Eq.~\eqref{3_10}, the radial magnetic field $H^a_r$ decreases asymptotically as $M/r^3$ (where $M$ is a constant).
This indicates that a distant observer does not see such an object like a color magnetic charge. On comparing the fall-off with Maxwell's electrodynamics, one might suppose that the solution obtained by us describes a non-Abelian magnetic dipole; therefore, it should be emphasised here that, in contrast to a Maxwellian dipole, the energy density of our system is spherically symmetric.
\item[(c)]
In the theory of superconductivity, there is a phenomenological description of such a phenomenon using the Ginzburg-Landau equation. On comparing the nonlinear Dirac equation and the Ginzburg-Landau equation, one can assume that, as it happens in the case of the Ginzburg-Landau equation, the normalization integral for the Dirac equation will determine the fermion number. In this case the nonlinear Dirac equation is not a fundamental equation but it is employed as some approximate approach to a phenomenological description of some quantum phenomena, as was suggested in Ref.~\cite{Dzhunushaliev:2019ham}.
\end{enumerate}

\section*{Acknowledgements}

The work was supported by the program No.~BR10965191 
(Complex Research in Nuclear and Radiation Physics, High Energy Physics and Cosmology for the Development of Competitive Technologies)
of the Ministry of Education and Science of the Republic of Kazakhstan.
We are also grateful to the Research Group Linkage Programme of the Alexander von Humboldt Foundation for the support of this research.
J.K. gratefully acknowledges support by the DFG Research Training Group 1620 {\sl Models of Gravity} and the COST Action CA16104.


\begin{thebibliography}{99}

\bibitem{Finkelstein:1951zz}
R.~Finkelstein, R.~LeLevier, and M.~Ruderman,
``Nonlinear Spinor Fields,''
Phys.\ Rev.\  {\bf 83}, 326 (1951).

\bibitem{Finkelstein:1956}
R. Finkelstein, C. Fronsdal, and P. Kaus,
``Nonlinear Spinor Field,''
Phys.\ Rev.\  {\bf 103}, 1571 (1956).

\bibitem{hei66} W. Heisenberg, {\it  Introduction to the unified field theory of elementary particles} (Max-Planck-Institut f\"ur Physik und Astrophysik, Interscience Publisher, London, 1966).

\bibitem{Nambu:1961tp}
Y.~Nambu and G.~Jona-Lasinio,
  ``Dynamical Model of Elementary Particles Based on an Analogy with Superconductivity.~I,''
  Phys.\ Rev.\  {\bf 122}, 345 (1961).

\bibitem{Nambu:1961fr}
Y.~Nambu and G.~Jona-Lasinio,
  ``Dynamical Model of Elementary Particles Based on an Analogy with Superconductivity.~II,''
  Phys.\ Rev.\  {\bf 124}, 246 (1961).

%\cite{Hatsuda:1994pi}
\bibitem{Hatsuda:1994pi}
T.~Hatsuda and T.~Kunihiro,
  ``QCD phenomenology based on a chiral effective Lagrangian,''
  Phys.\ Rept.\  {\bf 247}, 221 (1994).

%\cite{Dzhunushaliev:2020qwf}
\bibitem{Dzhunushaliev:2020qwf}
V.~Dzhunushaliev, V.~Folomeev, and A.~Serikbolova,
  ``Monopole solutions in SU(2) Yang-Mills-plus-massive-nonlinear-spinor-field theory,''
  Phys.\ Lett.\ B {\bf 806}, 135480 (2020).

\bibitem{Li:1982gf}
X.~z.~Li, K.~l.~Wang, and J.~z.~Zhang,
``Light Spinor Monopole,''
Nuovo Cim.\ A {\bf 75}, 87 (1983).

\bibitem{Li:1985gf}
K.~L.~Wang and J.~Z.~Zhang,
``The Problem of Existence for the Fermion-Dyon Selfconsistent Coupling System in a SU(2) Gauge Model,''
Nuovo Cim.\ A {\bf 86}, 32 (1985).

%\cite{Soler:1970xp}
\bibitem{Soler:1970xp}
 M.~Soler,
  ``Classical, stable, nonlinear spinor field with positive rest energy,''
  Phys.\ Rev.\ D {\bf 1}, 2766 (1970).

%\cite{Dzhunushaliev:2019ham}
\bibitem{Dzhunushaliev:2019ham}
V.~Dzhunushaliev, V.~Folomeev, and A.~Makhmudov,
  ``Non-Abelian Proca-Dirac-Higgs theory: Particlelike solutions and their energy spectrum,''
  Phys.\ Rev.\ D {\bf 99}, no. 7, 076009 (2019).
\end{thebibliography}
\end{document}